\documentclass{article}
\usepackage{spconf,amsmath,graphicx, amssymb,tabularx}
\usepackage{multirow}
\usepackage{enumitem}
\usepackage{multicol}
\usepackage{cite}
\usepackage[ruled, linesnumbered]{algorithm2e}
\usepackage[small,compact]{titlesec} 
\usepackage{times}
\usepackage[small,it]{caption} 
\usepackage{float}

\usepackage{comment}

\newcolumntype{D}{>{\centering\arraybackslash}m{6ex}}
\title{Accelerating RNN-T Training and Inference Using CTC guidance}
\name{
\begin{tabular}{c}
Yongqiang Wang, Zhehuai Chen, Chengjian Zheng, Yu Zhang, Wei Han, Parisa Haghani
\end{tabular}
  \thanks{We would like to acknowledge the invaluable discussions with our colleagues, Rohit Prabhavalkar, Ronny Huang, Kevin Hu, Weiran Wang, Bo Li, Shuo-yiin Chang,  Yanzhang He, Dongseong Hwang, Trevor Strohman and Pedro M. Mengibar.}
}
\address{Google}
\begin{document}
\ninept
\maketitle
\begin{abstract}\vspace{0.5em}
We propose a novel method to accelerate training and inference process of recurrent neural network transducer (RNN-T) based on the guidance from a co-trained connectionist temporal classification (CTC) model. We made a key assumption that if an encoder embedding frame is classified as a blank frame by the CTC model, it is likely that this frame will
be aligned to blank for all the partial alignments or hypotheses in RNN-T  and it can be discarded from the decoder input. We also show that this frame reduction operation can be applied in the middle of the encoder, which result in significant speed up for the training and inference in RNN-T. We further show that the CTC alignment, a by-product of the CTC decoder,  can also be used to perform lattice reduction for RNN-T during training. Our method is evaluated on the Librispeech and SpeechStew tasks. We demonstrate that the proposed method is able to accelerate the RNN-T inference by 2.2 times with similar or slightly better word error rates (WER).
\end{abstract}
\section{Introduction}
\label{sec:intro}

In recent years, end-to-end (E2E) modeling for automatic speech recognition (ASR) has been intensively studied and significant progress has been made (e.g. \cite{graves2013speech,chan2016listen, bahdanau2016end, soltau2016neural, chiu2018state, gulati2020conformer, karita2019comparative}). Broadly speaking, there are 3 different architectures under the E2E ASR category. Firstly, the connectionist temporal classification loss \cite{graves2006connectionist} can be used to optimize the likelihood of word or wordpiece \cite{schuster2012japanese} sequences (as compared to using phoneme sequences and finite state transducer in the traditional hybrid system, e.g., \cite{sak2015learning}). However, the lack of language modeling in this architecture usually leads to a sub-optimal recognition accuracy; Secondly, the attention-based sequence-to-sequence (S2S) modeling \cite{sutskever2014sequence, cho2014learning} can be adopted for E2E ASR, e.g. \cite{chan2016listen, bahdanau2016end}. However, this approach cannot naturally fit into the streaming requirements in many speech applications \cite{chiu2017monotonic}; the third approach, based on neural transducer loss \cite{graves2012sequence}, namely Recurrent Neural Network Transducer (RNN-T), integrates language models in the E2E model and fits well with the streaming requirement, therefore it  has been widely adopted \cite{he2019streaming}. Both RNN-T and CTC use a so-called ``blank" symbol to deal with the fact that the decoder input sequence is usually much longer than its output sequence. Notably, all these 3 architectures are equipped with an acoustic encoder which converts acoustic signals to a sequence of acoustic embeddings with a fixed frame rate.

Due to the volatile nature of speech signals, the encoder usually yields acoustic embeddings in a relatively high frequency. This results in that the decoder needs to process a much longer input sequence, stripping out information irrelevant to the ASR task, and output a much shorter sequence. In other words, there are considerable redundancy in the encoder output. There are many works in the past trying to leverage this redundancy. For example, \cite{sak2015fast} first proposed to use a frame rate larger than the 10ms in the traditional hybrid deep neural network (DNN) - hidden Markov model (HMM) system; \cite{Pundak2016} studied the effect of frame rate in DNN-HMM systems and found that such systems can work pretty well up to 60ms frame rate; based on the observation 
that the phonetic posteriorgram is usually dominated by blank frames, \cite{chen2016phone} proposed a Phone Synchronous Decoding (PSD) method for DNN-HMM systems and demonstrated that the best recognition hypothesis will not change when those blank-dominated frames are thrown away; this idea has been extended to RNN-T \cite{zhang2021tiny}, in which the search algorithm won't extend a partial hypothesis if its blank posterior probability is higher than some threshold; this has been further extended in \cite{tian2021fsr}, where during inference, a CTC model is first to calculate the posterior probability of the blank symbol -- if it is higher than some threshold, that frame will not be processed by the RNN-T decoder at all; \cite{zhang2020adaptive} proposed to select a small proportion of encoder output frames to feed to an attention-based sequence-to-sequence decoder in speech translation tasks.

\begin{figure}[tb]
    \centering
    \includegraphics[scale=0.042]{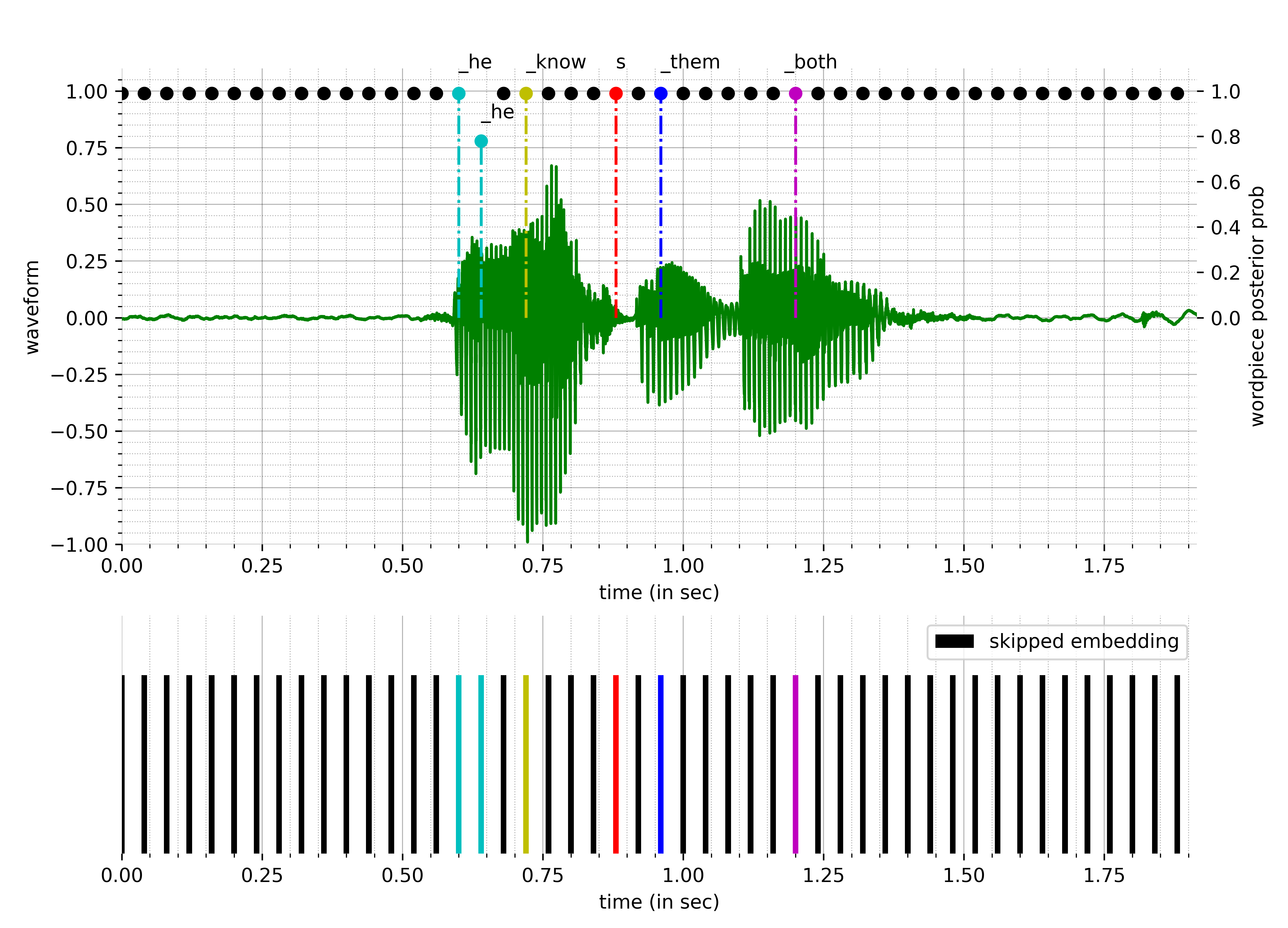}
    \caption{Illustration our main motivation using an utterance with \textit{he knows them both} as the transcription. The top figure shows the posterior probability provided by a CTC model overlaid with the actual audio signal; the bottom figure shows that all those blank symbol dominated encoder output embeddings are skipped by the decoder. }
    \vspace{0em}
    \label{fig:motivation}
\end{figure}

While there are different methods to compress the encoder output, we believe the posterior probabilities provided by a CTC model offer an intuitive way to select informative encoder output frames. We hypothesize that if a decoder output frame is determined by a CTC decoder that it is highly likely to be a blank frame, it is also very likely to become a blank frame for {\it all} RNN-T hypotheses, therefore we could skip this frame all-together in {\it both} training and inference. 
This is illustrated in Figure \ref{fig:motivation}. We further demonstrate that this frame reduction can happen in the middle of the encoder, which can accelerate both the encoder and the decoder. On the other hand, as a by-product of the CTC decoder, an alignment of the encoder frames with the reference label sequence is readily available. This alignment can be used to restrict the set of possible paths when calculating RNN-T loss, similar to \cite{mahadeokar2021alignment, sainath2020emitting} and \cite{kuang2022pruned}, where the first 2 works use external alignments obtained from another ASR system, while the latter uses an small RNN-T to obtain the alignment on-the-fly during training.
We validate our method on Librispeech (single domain) \cite{panayotov2015librispeech} and SpeechStew (multi-domain) \cite{chan2021speechstew} datasets.

\section{CTC and RNN-T}
\label{sec:ctc_rnnt}
In almost all the E2E ASR architectures, an acoustic encoder is used to process audio signals to produce a sequence of acoustic embeddings $\boldsymbol x_1, \cdots, \boldsymbol x_T$. The decoder process these acoustic embeddings to produce a label sequence $\boldsymbol l = (l_1, \cdots, l_U)$ and it is optimized by minimizing the following loss function: 
\begin{figure}
    \centering
    \includegraphics[scale=0.25]{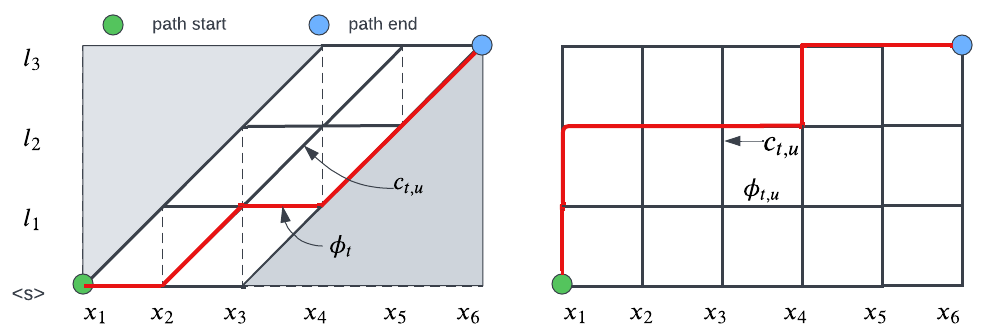}
    \caption{An comparison of lattices in CTC (left) and RNN-T(right). Bold lines are allowed paths; red lines are two example paths; gray areas are disallowed in the CTC lattice. $c_{t,u}$ is the probability of emitting the next symbol $l_{u+1}$ while sitting at position $(t,u)$, $\phi_{t,u}$ is the probability of emitting a blank symbol at the same place, whereas in CTC, $\phi_{t,u}$ does not depend on $u$, so it becomes $\phi_t$.}
    \label{fig:lattice}
\end{figure}
\begin{align}
    \mathcal{L}(\boldsymbol {l}) = -\log p(\boldsymbol{l} | \boldsymbol x_1, \cdots, \boldsymbol x_T)
\end{align}
Both CTC and RNN-T try to find all the possible alignment paths $\boldsymbol \pi$ between $\boldsymbol x_1, \cdots, \boldsymbol x_T$ and $\boldsymbol l$. Given a set of alignment paths that can be mapped to $\mathbf{l}$, $\mathcal{B}^{-1}(\mathbf l)$, the loss function can be rewritten as 
\begin{align}
    \mathcal L(\boldsymbol l) = - \sum_{\boldsymbol \pi \in \mathcal{B}^{-1}(\boldsymbol l)} \log p( \boldsymbol \pi| \boldsymbol x_1, \cdots, \boldsymbol x_T)
\end{align}
CTC further simplifies the loss function by making the following assumptions:
\begin{itemize}[leftmargin=*]
    \item At each time $t$, the decoder emits exactly one symbol, either one from $\boldsymbol l$ or a blank symbol $\phi$. Therefore, the alignment path $\boldsymbol \pi = (a_1, \cdots, a_T)$, where
    $a_t \in \mathcal{V} \cup \{\phi\}$, and $\mathcal{V}$ is the output vocabulary;
    \item label $l_u$ is conditionally independent of other symbols;
\end{itemize}
Using these assumptions, CTC loss function can be written as:
\begin{align}
    \mathcal L(\boldsymbol l) = - \sum_{\boldsymbol \pi \in \mathcal{B}^{-1}(\boldsymbol l)}\prod_{t=1}^{T}\log p(a_t | \boldsymbol x_t)
\end{align}
On the other hand, RNN-T relaxes these 2 assumptions that 
\begin{itemize}[leftmargin=*]
    \item At each time, the decoder is allowed to emit more than one symbol. Therefore the alignment path $\boldsymbol \pi = (\boldsymbol a_1, \cdots, \boldsymbol a_{T+U})$, where $\boldsymbol a_\tau = (t, u)$ indicates where the partial path ends;
    \item the probability of emitting next symbol $l_{u+1}$ or emitting the blank symbol $\phi$ also depends on $\boldsymbol l_{\leq u}$, the symbols emitted in the past.
\end{itemize}
Hence, the RNN-T loss function is written as
\begin{align}
    \mathcal L(\boldsymbol l) = - \sum_{\boldsymbol \pi \in \mathcal{B}^{-1}(\boldsymbol l)}\prod_{\tau=1}^{T+U}\log p(l_{u+1} | \boldsymbol x_t,  \boldsymbol{l}_{\leq u} )
\end{align}
In RNN-T, the probability distribution $p (\cdot | \boldsymbol x_t,  \boldsymbol{l}_{\leq u})$
is usually given by a joiner which takes $\boldsymbol x_t$ and the output of a predictor network as the input. The input to the predictor network is $\boldsymbol l_{\leq u}$.

In summary, both CTC and RNN-T are to minimize the overall cost of all the possible paths on a $T$-by-$U$ lattice. This can be illustrated from Figure \ref{fig:lattice} in which all the paths start from the bottom-left and end at top-right; at each position $(t,u)$, the decoder can choose to yield the next symbol $l_{u+1}$ with a cost $c_{t,u}$ or yield a blank symbol with a cost of $\phi_{t,u}$, whereas in CTC, the probability of yielding the blank symbol does not depend on the history, so $\phi_{t,u}$ becomes $\phi_t$ in CTC.

\begin{figure}
    \centering
    \includegraphics[scale=0.30]{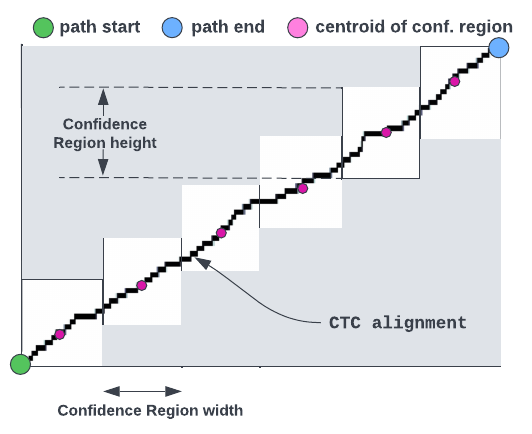}
    \caption{Illustration of lattice reduction based on CTC alignment.}
    \label{fig:ctc_lattice_redu}
\end{figure}

\vspace{-0.35em}
\section{RNN-T with CTC Guidance}
\begin{figure}[tb]
    \centering
    \includegraphics[scale=0.225]{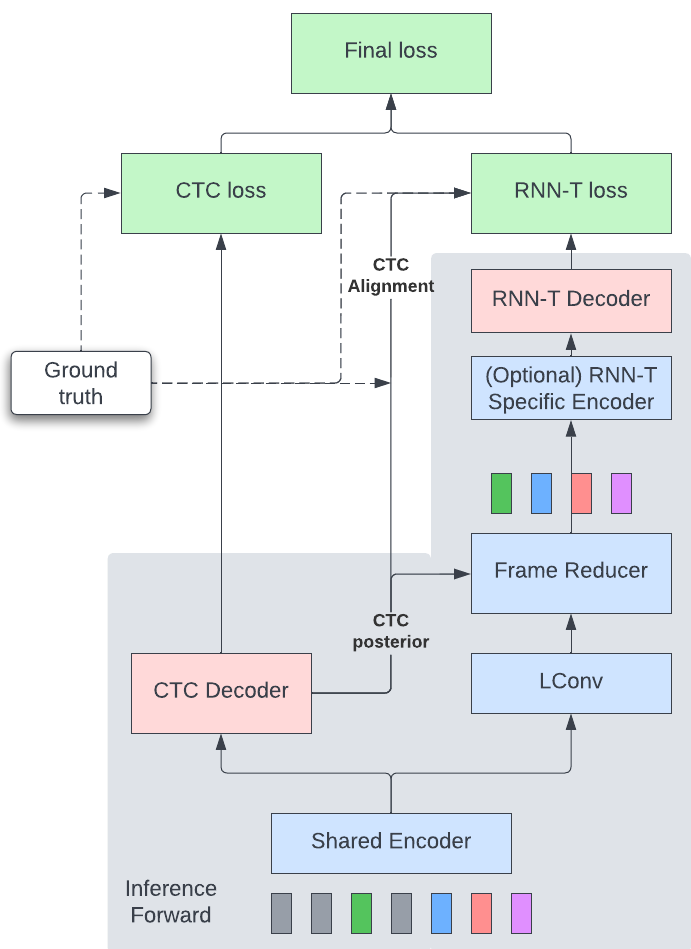}
    \caption{Overall architecture of the proposed model. Grey area shows the forward pass shared between training and inference. }
    \label{fig:overall}
\end{figure}

\vspace{-0.25em}
From Figure \ref{fig:lattice}, it can be seen that a complete path takes $T$ (CTC) or $T+U$ steps (RNN-T), while it can only emit $U$ symbols. In speech recognition, $T \gg U$, therefore both CTC or RNN-T alignment paths yield a lot of blank symbols. On the other hand, \cite{chen2016phone} points out that in CTC, since the cost of emitting a blank symbol at the $t$-th frame is independent of previous emitted symbols, if the blank symbol probability is high enough, there is no need to consider the $t$-th frame anymore, as it won't change the relative order of the partial hypotheses. However, this is not the case any more in RNN-T. In this work, we made a key assumption that if the $t$-th frame is likely to be a blank frame in CTC, it is also likely that all the partial paths in RNN-T will emit a blank symbol at time $t$. To this end, we propose a new RNN-T model, whereas it is multi-tasked trained with both CTC and RNN-T loss. During the forward pass, the encoder output is first used to calculate the CTC posterior probability; then for each output frame, if its blank posterior is bigger than some thresholds, it will be simply discarded from the encoder output. To prevent information loss, we also put a convolution module  similar to the one used in conformer\cite{gulati2020conformer} (referred to as ``LConv") before the frame reduction. We also notice that it is also possible to apply this frame reduction in the middle of the encoder, whereas only a few conformer layers are in the shared encoder, and rest of the layers can move to the RNN-T specific encoder, which process a much shorter sequence due to the frame reduction. The forward pass of our proposed model is illustrated in the grey area in Figure \ref{fig:overall}.

On the other hand, as shown in Figure \ref{fig:lattice}, RNN-T requires computing the probabilities of the all edges in the $T$-by-$U$ lattice. This needs to allocate a large amount of memory on graphic processing units (GPU) or tensor processor units (TPU). However, as pointed out in \cite{mahadeokar2021alignment} and \cite{kuang2022pruned}, not all alignment paths have high likelihoods, and most of the probability mass is assigned to the paths that are close to a reasonable alignment. As a by-product of the CTC decoder in our proposed system, we can easily get a CTC alignment by aligning the CTC posterior with the ground truth.  Then this CTC alignment can be used to construct a confidence region, and we restrict the RNN-T alignment paths must be within the confidence region. This is illustrated in Figure \ref{fig:ctc_lattice_redu}: we first divide the $T$-by-$U$ lattice into a few strips along the time axis; we then find the centroid of the CTC alignment with each strip, and construct a confidence region with a user specified height. Only the alignment paths that pass through the confidence region are considered valid, and the rest ones are simply pruned. In this way, the memory requirement and computation cost in calculating the RNN-T loss can be largely reduced.

\vspace{-0.5em}
\subsection{Relation to other works}
\vspace{-0.2em}
There are several previous works which proposed to use CTC to improve RNN-T systems. A notable work is \cite{tian2021fsr} where the encoder output embeddings are discarded based on the CTC posterior during the inference. However, those blank frames are still processed by the RNN-T decoder during training, resulting a mismatched between training and inference, thus degraded WERs when frame reduction is used. In \cite{liu2020modular} CTC-based frame reduction is also used to reduce the input frame to an attention-based sequence-to-sequence decoder, where their main motivation is to build modularized encoder and decoder. Sparsity of CTC posteriors are also studied and explored  in \cite{dalmia2022legonn} to improve the modularity. 
\vspace{-0.4em}
\section{Experiments}
\vspace{-0.3em}
\label{sec:exp}
We evaluate the effectiveness of the proposed method on both librispeech \cite{panayotov2015librispeech} (single domain) and 
SpeechStew \cite{chan2021speechstew} (multiple domains) tasks. Librispeech consisted of 960hrs read speech, 
while SpeechStew is a combination of various publicly available ASR datasets with a large variation of speaking styles and environment noises. 
We compare both recognition accuracy and efficiency of the proposed models with the standard RNN-T models.

\begin{table}[tb]
    \centering
    \begin{tabular}{|c|c|c||cc|cc|}
    \hline
    Mode & FR &  Decoder   &  \multicolumn{2}{c|}{WER}   &  \multicolumn{2}{c|}{RTF}  \\
         &         &         &  {\it other} & {\it clean} & Enc.  & Dec. \\
    \hline \hline
    \multirow{6}{*}{NS} & \multirow{2}{*}{--}  & CTC & 5.7 & 2.5 & 0.152 & 0.001 \\
                        &      & RNN-T & 4.4 & 2.0 & 0.152 & 0.066 \\
        \cline{2-7}
                                   & \multirow{2}{*}{Dec.} & CTC &  5.6 & 2.5  & 0.152 & 0.001\\
                                   &                       & RNN-T & 4.4 & 2.0  & 0.152 &  0.013\\
        \cline{2-7}
                                   & \multirow{2}{*}{Enc.} & CTC &   9.3 &  3.9  &  0.073  & 0.013\\
                                   &                        & RNN-T & 4.3 & 2.0   & 0.086  & 0.013 \\
    \hline\hline
    \multirow{5}{*}{S} & -- & RNN-T & 9.1 & 3.4 & 0.087 & 0.046 \\
        \cline{2-7}
                               & \multirow{2}{*}{Dec.} & CTC   & 11.9  & 4.7 & 0.087 & 0.001 \\
                               &                       & RNN-T & 9.1  & 3.5 & 0.087 &  0.023   \\
        \cline{2-7}
                               & \multirow{2}{*}{Enc.} & CTC & 16.1 & 6.6 & 0.045   & 0.001 \\
                               &                       & RNN-T &  9.1&  3.5 & 0.057 & 0.023\\
    \hline
    \end{tabular}
    \caption{WER and RTF comparisons of frame reduction on the librispeech task. ``NS" means ``non-streaming", ``S" means ``streaming"; ``FR`` indicates where the frame reduction is performed: ``--`` means no frame reduction is performed, ``Dec.`` means decoder frame reduction, ``Enc.`` means encoder frame reduction.}
    \label{tab:frame_reduc_ls}
\end{table}

In all of our experiments, a 128-dimensional log Mel-filter bank features are extracted from a 32ms window with a 10ms frame shift. These frames are then further processed by a convolution sub-sampling module with a total stride of 4. They are then fed to a 17-layer conformer encoder with 512 as the model dimension, similar to the ``ConformerL" in \cite{gulati2020conformer}. For streaming models, we mask all future frames in attention and use causal convolution in the encoder. Both the CTC and RNN-T decoder shared the same wordpiece model with a vocabulary size of 1024 (including blank symbol). The RNN-T decoder uses a 1-layer LSTM with hidden and memory cell dimension 640 as the predictor network. 

 The CTC and RNN-T losses are interpolated with a weight of 0.1 given to CTC and 1.0 to RNN-T. Frame reduction is performed when the blank posterior probability is larger than 0.9.  We found the ``LConv" module in Figure \ref{fig:overall} is critical to smooth the gradient after the frame reduction module. In this work, we borrow the convolution module in conformer \cite{gulati2020conformer}. This ``LConv" module contains a depthwise convolution with a kernel size of 7, sandwiched by 2 pointwise convolutions with an expansion factor of 2. 
Our training recipe follows \cite{gulati2020conformer}, where we use Adam optimizer \cite{kingma2014adam} with L2 regularization , variational noise \cite{graves2012sequence} and exponential moving average of model parameters. All the experiments in this work used 128 TPUv3 cores and each TPU core processing a batch of 16 utterances synchronously, resulting in a global batch size of 2048. When performing search, greedy decoding is used for CTC, and a beam size of 8 is used for RNN-T.
We measure the recognition accuracy using WERs. We performed inference benchmark on a EPYC 7B12 AMD CPU with 64 CPU cores and measure real time factors (RTF) of both encoders and decoders on the \textit{test-other} from the librispeech task. We did not use utterance batching during the inference. 
\vspace{-0.7em}

\begin{table*}[h]
    \centering
    \begin{tabular}{|c||cc|c|cc|cc|c|c|c|}
    \hline
    Model & \multicolumn{2}{c|}{AMI} & Common Voice & \multicolumn{2}{c|}{Librispeech} & \multicolumn{2}{c|}{SWB} & Ted-Lium & WSJ  & Chime-6\\
    \cline{2-10}
          & IHM & SDMI &   & clean & other & SWB & CH &  & eval92 &  \\
    \hline\hline
    Baseline \cite{chan2021speechstew} & 9.0 & 21.7 & 9.7 & 2.0 & 4.0 & 4.7 & 8.3 & 5.3 & 1.3 & 57.2 \\
    \hline
    Dec. FR                            & 9.5 & 22.3 & 10.0 & 2.1 & 4.2 & 4.7 & 9.1 & 6.2 & 1.4 & 55.4 \\
    Enc. FR                            & 9.1 & 21.2 & 9.3  & 1.9 & 3.9 & 4.7 & 9.0 &  5.9 & 1.2    &  52.2    \\
    \hline
    \end{tabular}
    \caption{WER comparisons of frame reduction on the multi-domain SpeechStew task.}
    \vspace{-2em}
    \label{tab:speechstew}
\end{table*}
\vspace{-0.2em}
\subsection{Results}
\vspace{-0.2em}
For the first set of experiment, we evaluate our proposed method on the librispeech task. Both ``non-streaming" and ``streaming" mode are considered, whereas in the streaming mode, both the shared encoder and RNN-T specific encoder produce acoustic embedding based on the partial input sequence up to the current time step. It is known that streaming mode usually produce a sub-optimal WER due to the limit access to acoustic context; it also causes an potential challenge to our proposed model, where CTC makes the frame reduction decision based only on the current frame plus the left context. As shown in Figure \ref{fig:overall}, frame reduction can be applied in different points. We apply the frame reduction 1) after all 17 conformer layers, refereed to as ``Decoder Frame Reduction"; 2) after 7 conformer layers, and the rest 10 conformer layers are moved to ``RNN-T specific encoder" in Figure \ref{fig:overall}, referred to as ``Encoder Frame Reduction". Since we have 2 decoders, we can switch to any of them as our main decoder during inference. We present WERs from both decoders in Table \ref{tab:frame_reduc_ls} though our primary goal is to focus on RNN-T decoder.

Experimental results under ``non-streaming" mode are listed in the first section in Table \ref{tab:frame_reduc_ls}. Our baseline is a co-trained  CTC/RNN-T system with a shared encoder but no frame reduction is performed.
This baseline achieves WERs of 4.4/2.0 on \textit{test-other} and \textit{test-clean} respectively. The same RNN-T system trained without CTC loss achieves the same WERs on these test sets, showing that adding a CTC loss does not hurt nor improve the RNN-T system. When we apply encoder or decoder frame reduction, we observed that around 72-75\% of the frames are reduced during training. Similar amount of frames are reduced on those test sets as well. Unfortunately, due to the static shape requirement in TPU, we cannot accelerate RNN-T training using frame reduction since it will generate a dynamic shape and we have to use paddings to satisfy the static shape requirement. Our frame reduction method should yield considerable speed up for training for other accelerating devices such as GPU. 
In terms of recognition accuracy, we did not see degradation when frame reduction is used, and encoder frame reduction achieve slightly better WERs, even though the CTC decoder got much worse WERs on two test sets (9.3\%/3.9\%) due to the fact the CTC decoder used the acoustic embedings from a relative shallow encoder (only 7 conformer layers).
Regarding the inference RTF, as expected the CTC greedy decoder is  cheap (0.001x RTF) compared with the RNN-T decoder (0.066x). With the decoder frame reduction, we saw a large RTF improvement of the RNN-T decoder, from 0.066x to 0.013x. This is in line with our observation that about 75\% of the frames are removed and not fed to the RNN-T decoder at all. When ``Encoder frame reduction" is used, we observe RTF improvements from both encoder and decoder, since a large portion of the encoder (10 out of 17 conformers) operates at a 4-5 times higher frame rate. This results in an 
encoder RTF improvement from 0.152 to 0.086. Overall, encoder frame reduction yields 2.2 times speed up compared with the standard RNN-T with no loss of recognition accuracy. 

Experimental results under ``streaming" mode are listed in the second section of Table \ref{tab:frame_reduc_ls}. It is within our expectation that WERs are worse than ``non-streaming" mode, particular for the CTC decoder. However, we again did not see noticeable recognition accuracy drop when CTC-based frame reduction is used. This demonstrates that our proposed method is relatively insensitive to the quality of the CTC decoder, as it is only used for blank/non-blank classification. On the other hand, we do observe significant RTF improvement when frame reduction is used. We also note that the decoder speed up in the streaming mode (0.046 to 0.023) is less than the speed up in the non-streaming model (0.066 to 0.013), this is because the CTC decoder has no access to the right context, therefore making less confident (probability $> 0.9$) classification about blank frames.

Librispeech is a read-speech task with a relatively low WER. To evaluate our method's robustness under various environments and application scenarios, we compare our method with the baseline RNN-T system on the multi-domain SpeechStew task. We follow the recipe in \cite{chan2021speechstew} in which training data from corpus like AMI (100hrs), Common Voice (1,500hrs), Librispeech (960hrs), Switchboard (SWB) and Fisher (2,000hrs), Ted-LIUM (450hrs), and Wall Street Journal (WSJ) (80hrs) are mixed together. The test sets from these corpus contains various speech styles and environment noise, reverberation etc. Chime-6 test set is used as a surprise domain test set, as its training data is not seen during training. We compare our method with the baseline RNN-T system (no CTC loss this time) in Table \ref{tab:speechstew}. We observed a slight degradation of decoder frame reduction method while encoder frame reduction achieve similar WERs as the baseline. The inference RTF benchmark is similar to what we observed in Table \ref{tab:frame_reduc_ls}.

\begin{table}[htbp]
\vspace{0em}
    \centering
    \begin{tabular}{|cc|cc|cc|}
    \hline
    \multicolumn{2}{|c|}{Conf. Region} & \multicolumn{2}{c|}{Librispeech WER} &  \multicolumn{2}{c|}{Dec. Time}  \\
    width & height             & \textit{other} & \textit{clean}            &  CTC & RNN-T                     \\
    \hline\hline
    8 & $\infty$      & 4.5   & 2.1   & -- & 280ms \\
    \hline
       8 &  33      & 4.5   & 2.1   & \multirow{2}{*}{4ms} &  124ms \\
       8 &      17      & 4.6   & 2.1   &     & 103ms \\
    \hline
    \end{tabular}
    \caption{Effect of lattice reduction based on CTC alignment. ``Dec. Time" means the TPU time spent on decoders during training for one batch.}
    \label{tab:lattice_reduction}
    \vspace{-0.2em}
\end{table}

In the last set of experiments, we combined the decoder frame reduction with lattice reduction. Note that, different from the RNN-T loss implementation in \cite{yang2022torchaudio} which calculates the costs of all the edges in the RNN-T lattice in one pass, our baseline implementation already splits the lattice along the time axis into several strips (strip length 8 in our experiments, i.e., the confidence region width equal to 8 in Figure \ref{fig:ctc_lattice_redu}) and calculates the edge cost in each strips sequentially. This implementation avoids allocating huge TPU memory, but it is relatively slow because of the sequential computation which is not friendly to TPU. On top of this baseline implementation, we construct confidence regions with a user-specified confidence region height around the CTC alignment, and only calculate the cost of edges within these confidence regions. Compared with our baseline implementation, we mainly save computation. 
On the other hand, our proposed lattice reduction method can equally applied to other implementation like the one in \cite{yang2022torchaudio} and could save both computation and TPU/GPU memory considerably. Results are presented in Table \ref{tab:lattice_reduction}. We observe that in the baseline implementation, about 280ms is spent on RNN-T decoder, including predictor and joiner networks, RNN-T loss calculation and the back-propagation through these components. To calculate the CTC alignment, we spent an extra 4ms. In return, with the CTC alignment, we save more by reducing the computation on the RNN-T lattice. With a confidence region height of 33, the time spent on computing RNN-T decoder per step is reduced to 124ms with no loss of recognition accuracy \footnote{Table \ref{tab:lattice_reduction} uses the same model and task as in Table \ref{tab:frame_reduc_ls}. However, the joiner implementations are slightly different, thus resulting in a 0.1 WER difference in the baseline WERs. }. With a confidence region of 17, the decoder time is further reduced to 103ms, a 2.7x speed up compared to the baseline. 
\vspace{-0.2em}

\section{Conclusions and Discussions}
\vspace{-0.2em}

We present a novel method to accelerate training and inference for RNN-T models based on the CTC guidance.
By using a cheap CTC decoder, we perform both frame reduction and lattice reduction for the relative expensive RNN-T. We also show that the frame reduction can be applied much earlier resulting in a significant RTF improvement in both encoder and decoder with no loss of recognition accuracy. 

One interesting observation from our experimental results is that the CTC-based encoder frame reduction achieves  better results than the decoder frame reduction. A possible explanation is that after blank frames are removed, the conformer encoders can focus on informative frames which represents more like text tokens; therefore performs more like language modeling. This could enable us to learn from speech and text modalities simultaneously. We will explore this direction in our future work. 

\label{sec:con}

\clearpage
\bibliographystyle{IEEEbib}
\bibliography{strings,refs}

\end{document}